\documentclass[12pt,a4paper]{article}
\usepackage[T1]{fontenc} 
\usepackage[dvipdfmx]{graphicx}
\usepackage{bm,caption,latexsym,amssymb,amsfonts,mathtools,here}
\usepackage{bm,latexsym,amsfonts,mathtools}
\usepackage{amssymb, amsmath, comment}
\usepackage{color}
\usepackage{hyperref}
\usepackage{booktabs}
\usepackage{mathrsfs}
\usepackage{wrapfig}
\usepackage{setspace}
\numberwithin{equation}{section}
\renewcommand{\baselinestretch}{1.5}

\usepackage[sorting=none,doi=false]{biblatex}

\bibliography{reference}

\begin{document}
\begin{titlepage}
\unitlength = 1mm
\begin{flushright}
KOBE-COSMO-24-01
\end{flushright}

\vskip 1cm
\begin{center}

{ \large \textbf{Chromo-natural warm inflation}}

\vspace{1.8cm}
Ann Mukuno$^\flat$ and Jiro Soda$^{\flat,\sharp}$

\vspace{1cm}

\shortstack[l]
{\it $^\flat$ Department of Physics, Kobe University, Kobe 657-8501, Japan\\
\it $^\sharp$ International Center for Quantum-field Measurement Systems for Studies
of the Universe and\\ 
\ \  \it  Particles (QUP), KEK, Tsukuba 305-0801, Japan
}

\vskip 4.0cm

{\large Abstract}\\
\end{center}
Chromo-natural inflation is a model where non-abelian gauge fields are sustained by the coupling of the axion with the gauge field through the Chern-Simons term. While minimal warm inflation is a model where the axion produces a thermal bath of non-abelian gauge particles through the Chern-Simons term. Since both axion inflation models are based on the same action, a natural question is if those are compatible or not. We study axion inflation with the Chern-Simons term and find that chromo-natural inflation can accommodate radiation with a temperature much larger than the Hubble parameter during inflation, which is a characteristic feature of warm inflation.
Thus, we conclude that chromo-natural warm inflation exists, which must have phenomenologically interesting consequences.

\vspace{1.0cm}
\end{titlepage}

\hrule height 0.075mm depth 0.075mm width 165mm
\tableofcontents
\vspace{1.0cm}
\hrule height 0.075mm depth 0.075mm width 165mm
\section{Introduction}
  The inflationary scenario can be realized by a single scalar field, so-called inflaton. In the slow-roll
  limit, the inflaton acts as a cosmological constant. The cosmic no-hair theorem~\cite{Wald:1983ky} tells us that the exponential expansion of the universe driven by a cosmological constant erases all of the initial information of the universe. More precisely,  in the presence of the cosmological constant, the following occurs 
\begin{itemize}
\item[(1)] The energy density of ordinary matter vanishes. 
\item[(2)] The anisotropy of spacetime vanishes. 
\item[(3)] The spatial curvature (except for the Bianchi IX spacetime with a very large curvature scale compared to the Hubble scale) goes to zero. 
\end{itemize}
Consequently,  the universe becomes vacuum with homogeneous and isotropic flat spacetime where only quantum vacuum fluctuations remain. 
This is the reason why inflation predicts the scale invariant spectrum with a statistically isotropic Gaussian distribution~\cite{Soda:2012zm}.  
The belief that any inflation model leads to this prediction is named the cosmic no-hair conjecture.

In the presence of additional fields during inflation, however, conventional inflation could be unstable and a novel inflationary phase with cosmic hairs could appear.  In fact, there exist two clear counterexamples to the cosmic no-hair conjecture, namely, warm inflation~\cite{Berera:1995ie} which violates (1) and anisotropic inflation~\cite{Watanabe:2009ct} assisted by gauge fields which violates (1) and (2).  Generically, those novel inflation models give rise to qualitatively new predictions  such as the statistically anisotropic non-Gaussianity~\cite{Yokoyama:2008xw}.

From the particle physics point of view, it is important to explore qualitatively novel inflation models with gauge fields~\cite{Maleknejad:2012fw}. 
As to the case of abelian gauge fields, 
it has been shown that anisotropic inflation and warm inflation  yield anisotropic warm inflation~\cite{Kanno:2022flo}.
In the case of non-abelian gauge fields,
chromo-natural inflation~\cite{Peter:2012ch,Peter:2012ch2} 
and minimal warm inflation~\cite{Berghaus:2019whh,DeRocco:2021rzv}   are the counter examples to the cosmic no-hair conjecture.  In chromo-natural inflation, the background gauge field exists and interacts with the axion through the Chern-Simons term to achieve slow-roll inflation. 
 In minimal warm inflation, on the other hand, the axion interacting with the gauge field through Chern-Simons term produces a thermal bath, which induces a friction term in the axion dynamics realizing slow-roll inflation. Since both inflation models utilize the similar setup, it is intriguing to study if chromo-natural inflation and warm inflation can coexist or not.

Curiously, although both models have the same action
including Chern-Simons interaction term,  the predictions of both models seem apparently different. In fact, chromo-natural inflation predicts the excess of primordial gravitational waves due to the instability of gauge field perturbations~\cite{Dimastrogiovanni:2012ew,Peter:2013,Adshead:2013qp,Namba:2013kia}. 
While, in minimal warm inflation, it is a challenge to have detectable primordial gravitational waves~\cite{Berghaus:2019whh,Klose:2022rxh}. It is natural to ask which happens in reality. Logically, both inflation may occur simultaneously, namely, chromo-natural warm inflation may exist. 
In that case, it is interesting to reveal the prediction of the model on primordial gravitational waves. Thus, it is worth investigating the phase space structure of axion inflation
with the non-abelian Chern-Simons term.

The purpose of this paper is to clarify chromo-natural warm inflation occurs or not.
We study the effect of dissipation of inflaton energy on chromo-natural inflation by solving the slow roll equation analytically and numerically.
When the effect of dissipation is negligible, chromo-natural inflation occurs.
 If we increase the dissipation rate, chromo-natural inflation becomes warm. 
 Thus, it turns out that chromo-natural inflation could accommodate a thermal bath with a temperature $T$ much larger than the Hubble parameter $H$, which is a characteristic feature of warm inflation.
 This is nothing but chromo-natural warm inflation. If we further increase the dissipation rate, chromo-natural inflation disappears and only warm inflation occurs. Eventually, for very large dissipation, only cold inflation remains.

The paper is organized as follows. In section 2, we review chromo-natural inflation. In section 3, we review minimal warm inflation. In section 4, we present the slow-roll equations of natural inflation with the Chern-Simons term and clarify the slow roll conditions. First, we analytically investigate the slow roll equations and verify the existence of chromo-natural warm inflation.  Next, we study the slow roll equations semi-analytically and reveal the phase space structure of axion inflation.
As a consequence, we prove the compatibility of chromo-natural inflation and warm inflation. 
The final section is devoted to conclusion.

\section{A review of chromo-natural inflation}
\label{section2}
In this section, we review chromo-natural inflation where the axion is coupled with non-abelian gauge fields through the Chern-Simons term~\cite{Peter:2012ch, Peter:2012ch2}. 
The interaction between the gauge field and the axion leads to an extra friction which realize the slow-roll inflation
even for the steep potential. Hereafter, we shall work in natural units $(c=\hbar=M_{pl}=1)$.

The chromo-natural inflation is described by the action
\begin{eqnarray}
S=\int d^4 x \sqrt{-G}
\left[
\frac{1}{2} R -\frac{1}{2} \partial_\mu \chi \partial^\mu \chi -V(\chi)-\frac{1}{4}F^a_{\mu\nu}F_a^{\mu\nu}-\frac{g^2 \lambda}{8 f}\chi \epsilon^{\mu\nu\lambda\rho}F^a_{\mu\nu} F^a_{\lambda\rho}
\right]\,,
\label{action1}
\end{eqnarray}
where $G$ is the determinant of the metric $G_{\mu\nu}$, $R$ is the scalar curvature, $\chi$ is an axion field , and the field strength of non-abelian gauge field is $F^a_{\mu\nu}=\partial_\mu A_\nu^a- \partial_\nu A_\mu^a\ +g\epsilon^{abc}A_\mu^b A_\nu^c$, and $g$ is a coupling constant of the non-abelian gauge field.
Greek letters denote spacetime indices, while Roman letters denote gauge indices. The last term is the Chern-Simons term with a coupling
constant $\lambda$. We took the convention $\sqrt{-G} \ \epsilon^{0123}=1$. The axion potential $V(\chi)$ reads 
\begin{eqnarray}
V(\chi)=\mu^4\left(1+\cos{\frac{\chi}{f}}\right)
\label{potential}
\end{eqnarray}
where the energy scale is given by  $\mu$ and
$f$ represents the decay constant.

Let us consider the cosmological background solution
with the metric
\begin{eqnarray}
ds^2=-dt^2+a(t)^2\,\delta_{ij}\,dx^idx^j\,,
\end{eqnarray}
where $a(t)=e^{\alpha(t)}$ is a scale factor and  $H=\dot\alpha$
is the Hubble parameter. 
The gauge field configuration is taken to be homogeneous and isotropic $A^a_i=\psi(t) a(t) \delta^a_i$ and $A^a_0=0$. 
 Then, Einstein equations are given by  
\begin{align}
&\dot\alpha^2=\frac{1}{3}\left[\frac{1}{2}\dot\chi^2+\mu^4\left(1+\cos{\frac{\chi}{f}}\right)+\frac{3}{2}\dot\psi^2+3\dot\alpha\dot\psi\psi+\frac{3}{2}\dot\alpha^2\psi^2+\frac{3}{2}g^2\psi^4 \right] \, ,
\label{EOMCNI1}\\
&\ddot\alpha=-\frac{1}{2}\dot\chi^2-\dot\psi^2 -2\dot\alpha\dot\psi\psi
-\dot\alpha^2\psi^2-g^2\psi^4\, .
\label{EOMCNI2}
\end{align}
The field equations for the axion and the gauge field can be deduced as
\begin{align}
&\ddot\chi+3\dot\alpha \dot\chi-\frac{\mu^4}{f}\sin{\frac{\chi}{f}}=-3\frac{g^3\lambda}{f}\psi^2(\dot\psi+\dot\alpha \psi) \, ,
\label{EOMCNI3}\\
&\ddot\psi+3\dot\alpha\dot\psi+\left[-\frac{1}{6}\dot\chi^2+\frac{2}{3}\mu^4\left(1+\cos{\frac{\chi}{f}}\right)\right]\psi+2g^2\psi^3-\frac{g^3 \lambda}{f}\psi^2\dot\chi=0 \, .
\label{EOMCNI4}
\end{align}
In the slow-roll approximation where we impose the condition $|\ddot\chi|\ll \dot\alpha|\dot\chi|$, $|\ddot\psi|\ll \dot\alpha|\dot\psi|$ and the inflaton potential is dominant in Eq.~\eqref{EOMCNI1}, we need to solve equations 
\begin{align}
&\dot\alpha^2=\frac{\mu^4}{3}\left(1+\cos{\frac{\chi}{f}}\right) \, ,
\label{EOMCNIs1}\\
&3\dot\alpha \dot\chi-\frac{\mu^4}{f}\sin{\frac{\chi}{f}}=-3\frac{g^3\lambda}{f}\psi^2(\dot\psi+\dot\alpha \psi) \, ,
\label{EOMCNIs3}\\
&3\dot\alpha\dot\psi+\frac{2}{3}\mu^4\left(1+\cos{\frac{\chi}{f}}\right)\psi+2g^2\psi^3-\frac{g^3\lambda}{f}\psi^2\dot\chi=0 \, .
\label{EOMCNIs4}
\end{align}
For $\lambda\ne0$, the consistency conditions for the slow-roll approximation
read
\begin{align}
&\epsilon_{\rm v}\equiv\frac{1}{2\left(1+\frac{\sigma^2}{3}\right)}\left(\frac{V'(\chi)}{V(\chi)}\right)^2\ll 1 \,, \label{cnsc1}\\
&\eta_{\rm v}\equiv\frac{1}{\left(1+\frac{\sigma^2}{3}\right)}\frac{V''(\chi)}{V(\chi)}\ll 1 \,,\label{cnsc2}
\end{align}
and
\begin{align}
\delta\equiv\frac{1}{\left(1+\frac{\sigma^2}{3}\right)}\frac{\sigma^2}{\gamma}\ll 1\label{cnsc3} \, ,
\end{align}
where we defined $\sigma\equiv\frac{g^3\lambda\psi^2}{f\dot\alpha}$ and $\gamma\equiv\frac{g^4\lambda^2\psi^2}{f^2}$. Notice that the slow roll parameters have the extra factor $1+\sigma^2/3 $. 
For $\lambda=0$, we simply obtain the conventional slow-roll conditions of natural inflation. 
\begin{align}
&\epsilon_{\rm v}=\frac{1}{2}\left(\frac{V'(\chi)}{V(\chi)}\right)^2\sim\frac{1}{2f^2}\tan{^2\frac{\chi}{2f}}\ll1  \,,
\label{NIe}\\
&\eta_{\rm v}=\frac{V''(\chi)}{V(\chi)}\sim\frac{1}{2f^2}\left(1-\tan{^2\frac{\chi}{2f}}\right)\ll1 \, .
\label{NIh}
\end{align}
We see that inflation can occur only for  $f \gg 1$ in Planck units, which is unnatural from a particle physics perspective. 
For $\lambda\ne 0$, however,  $f$ can take a natural value,
if $\sigma\gg1$ and $\gamma\gg1$. The conditions $\sigma \gg 1$ and $\gamma\gg1$ require 
$\frac{g^2\lambda}{f} \gg 1$, because we need $g^2 \psi^4 \ll \mu^4\sim \dot\alpha^2$ 
and $\psi \ll 1$ for inflaton potential to be dominant in \eqref{EOMCNI1}.

We rewrite Eqs.~\eqref{EOMCNIs3} and \eqref{EOMCNIs4} using $\sigma$ and $\gamma$ as
\begin{align}
&3\dot\alpha \dot\chi=\frac{\mu^4}{f}\sin{\frac{\chi}{f}}-3\sigma\dot\alpha(\dot\psi+\dot\alpha \psi) ~ ,
\label{EOMCNIs5}\\
&3\dot\alpha\dot\psi=-2\left(1+\frac{\sigma^2}{\gamma}\right)\dot\alpha^2\psi+\sigma\dot\alpha\dot\chi ~ .
\label{EOMCNIs6}
\end{align}
From Eqs.~\eqref{EOMCNIs5} and \eqref{EOMCNIs6}, we can deduce the equations for $\dot\psi$ and $\dot\chi$. By imposing the slow-roll conditions, $\sigma\gg1$ and $\gamma\gg1$, we obtain the equation for the gauge field
\begin{align}
&\dot\alpha\dot\psi\simeq\frac{\frac{\mu^4}{f}\sin{\frac{\chi}{f}}}{3\sigma}-\dot\alpha^2\psi \, .
\label{EOMCNIs3-1}
\end{align}
Defining an effective potential for the gauge field,
\begin{align}
V_{\rm eff}(\psi)=\dot\alpha^2\frac{\psi^2}{2}+\frac{\mu^4\sin{\frac{\chi}{f}}}{3g^3\lambda}\frac{\dot\alpha}{\psi} \, ,
\label{Veff}
\end{align}
we can rewrite Eq.~\eqref{EOMCNIs3-1} as 
\begin{align}
&\dot\alpha\dot\psi\simeq-V_{\rm eff}' \, .
\label{EOMCNIs1-3}
\end{align}
Thus, we see the minimum of the effective potential  
\begin{align}
&\psi_{\rm min}\simeq \left(\frac{\mu^4\sin{\frac{\chi}{f}}}{3g^3\lambda \dot\alpha}\right)^{1/3} \, 
\label{psimin}
\end{align}
is an attractor. Indeed, from $V_{\rm eff}$, the effective mass of the gauge field $\psi$ at $\psi_{\rm min}$ can be calculated as $m_\psi^2\simeq 3\dot\alpha^2$, which is positive and large.
Hence, the minimum is an attractor.  
Now, the slow-roll conditions can be expressed as 
\begin{align}
&\sigma\simeq\frac{g^3\lambda\psi_{\rm min}^2}{f\dot\alpha}=\left(\frac{g^3\lambda\mu^8\sin^2{\frac{\chi}{f}}}{3^2 f^3\dot\alpha^5}\right)^{\frac{1}{3}}\gg 1 \, ,
\label{sigmapsimin}\\
&\gamma\simeq\frac{g^4\lambda^2\psi_{\rm min}^2}{f^2}=\left(\frac{g^6\lambda^4\mu^8\sin^2{\frac{\chi}{f}}}{3^2 f^6 \dot\alpha^2}\right)^{\frac{1}{3}}\gg 1 \,.
\label{gammapsimin}
\end{align}

Now, the equation of axion reads from Eq.~\eqref{EOMCNIs6} by taking $\dot\psi\simeq0$
\begin{align}
&\dot\alpha\dot\chi\simeq\frac{2}{\sigma}\left(1+\frac{\sigma^2}{\gamma}\right)\dot\alpha^2\psi  \,.
\label{EOMCNIs4-1}
\end{align}
Using Eqs.~\eqref{EOMCNIs1},~\eqref{psimin} and \eqref{EOMCNIs4-1}, 
we can calculate the number of e-folding as
\begin{eqnarray}
N=\int{\frac{\dot\alpha}{\dot\chi}~d\chi}\simeq \int{\frac{\sigma}{2\left(1+\frac{\sigma^2}{\gamma}\right)\psi}}~d\chi=g^2\lambda\int^{\pi}_{\frac{\chi_i}{f}} {\frac{\beta(1+\cos{X})^{\frac{2}{3}}\sin^{\frac{1}{3}}{X}}{3^{-\frac{1}{3}}\beta^2(1+\cos{X})^{\frac{4}{3}}+\sin{X}^{\frac{2}{3}}}}~dX\,,
\label{ec}
\end{eqnarray}
where we defined $X=\chi/f$ and $\beta=\lambda^{1/3}\mu^{4/3}$. For the current parameters, $\sigma \gg 1$ and $\gamma \gg 1$ (for example, $g\sim 10^{-6}$, $g^2\lambda\sim 10^2 $ and $\mu\sim10^{-4}$), we have $\beta\sim\mathcal{O}(1)$. If we take $X=10^{-2}$ as an initial value, the number of e-folding is given by  $N\sim g^2\lambda~\mathcal{O}(1)$. Therefore, $g^2\lambda\sim \mathcal{O}(100)$ is enough for sufficient inflation to occur.

\section{A review of minimal warm inflation}
\label{section3}
 Warm inflation is an attractive model because it might provide a mechanism for slow-roll inflation without a shallow potential and thermalization of the universe without reheating~\cite{Berera:1995ie,Berera:1995wh,Berera:2008ar}. The point is that the decay rate of the inflaton satisfies $\Gamma_\chi \gtrsim H$ during inflation. The inflaton produces other light fields  and generates a thermal bath. However, since the thermal bath gives a temperature typically much larger than the Hubble parameter, thermal back-reaction to the inflaton potential may destroy the warm inflation scenario~\cite{Yokoyama:1998ju}.  To resolve the thermal backreaction issue, minimal warm inflation utilizing the shift symmetry of an axion has been proposed~\cite{Berghaus:2019whh}. Although there are other models protecting quantum and thermal corrections to the inflaton potential~\cite{Berera:1998px,Berera:2003kg,Bastero:2016qru,Bastero:2019gao}, we focus on the minimal warm inflation model in this paper. 

The Lagrangian of minimal warm inflation
 driven by the axion field $\chi$ is given by
\begin{eqnarray}
S=\int d^4 x \sqrt{-G}
\left[
\frac{1}{2} R -\frac{1}{2} \partial_\mu \chi \partial^\mu \chi -V(\chi) 
 -\frac{1}{4}F^a_{\mu\nu}F_a^{\mu\nu}-\frac{g^2 \lambda}{8 f}\chi \epsilon^{\mu\nu\lambda\rho}F^a_{\mu\nu} F^a_{\lambda\rho} 
\right]\,.
\label{action2}
\end{eqnarray}
The action is exactly the same as that of chromo-natural inflation. 
In the scenario of warm inflation, it is crucial to take into account the dissipation rate of the inflaton $\Gamma_\chi$ calculated from the sphaleron transition rate~\cite{Peter:1987} as
\begin{eqnarray}
\Gamma_\chi(T)=\kappa g^{10} T^3\frac{\lambda^2}{f^2}
\,,
\label{Gamma}
\end{eqnarray}
where the parameter $\kappa$ depends on $g$, the number of colors $N_c$ and flavors $N_f$ of the gauge group~\cite{Moore:2011sp}. The natural value is $\kappa \sim \mathcal{O}(100)$.
It is convenient to define the dimensionless quantity
$Q$ by the relation $\Gamma_\chi=3HQ$, namely, 
\begin{eqnarray}
Q =\kappa g^{10} T^3\frac{\lambda^2}{3H f^2}\,.
\label{dissipation}
\end{eqnarray}
In warm inflation, the dissipation of the energy of the axion field $\chi$ due to the coupling to gauge fields leads to an effective friction in the equation of the axion
\begin{eqnarray}
\ddot{\chi}+3H\left(1+Q\right)\dot{\chi}+V'\left(\chi\right)=0 \, .
\label{eom1}
\end{eqnarray}
Moreover, the Friedman equation reads
\begin{eqnarray}
H^2=\frac{1}{3M_{\rm pl}^2} \left(V(\chi)+\frac{1}{2}\dot{\chi}^2+\rho_{\rm R}\right) \,,
\label{eom2}
\end{eqnarray}
where we took into account the energy density of radiation 
with the internal degrees of freedom $g_i$
\begin{eqnarray}
   \rho_R = \frac{\pi^2}{30}\ g_i T^4 \label{rho} \, . 
\end{eqnarray}
The conservation of energy is described by
\begin{eqnarray}
\dot{\rho}_{\rm R}+4H\rho_{\rm R}=3H Q \dot{\chi}^2 \, .
\label{eom3}
\end{eqnarray}
Using the slow-roll approximation $|\ddot{\chi}|\ll H\,|\dot{\chi}|$ in Eq.~\eqref{eom1},  $\dot{\chi}^2\ll V(\chi)$,   $\rho_{\rm R}\ll V(\chi)$ in Eq.~\eqref{eom2} and $\dot\rho_{\rm R}\ll H\rho_{\rm R}$ in Eq.~\eqref{eom3}, we obtain
the slow-roll equations
\begin{eqnarray}
&&3H(1+Q)\,\dot{\chi}+V'(\chi)=0 \, ,
\label{EOM1}\\
&&H^2=\frac{V(\chi)}{3M_{\rm pl}^2} \, ,
\label{EOM2}\\
&&\rho_{\rm R}=\frac{3}{4} Q \dot{\chi}^2\, .
\label{EOM3}
\end{eqnarray}
The consistency conditions for the slow-roll approximation are given by
\begin{align}
&\epsilon_{\rm v}\equiv\frac{1}{2(1+Q)}\left(\frac{V'(\chi)}{V(\chi)}\right)^2 \ll1 \,, 
    \label{sc4}\\
&\eta_{\rm v}\equiv\frac{1}{1+Q}\frac{V''(\chi)}{V(\chi)}\ll 1 \,,\label{sc5}\\
&\beta\equiv\frac{1}{1+Q}\frac{Q'\,V'(\chi)}{Q\,V(\chi)}\ll 1 \, .
\label{sc6}
\end{align}
We can see that the factor $1+Q$ alleviates the condition on the inflaton potential.

Since we know the axion potential \eqref{potential}, we can calculate the number of  e-folding as
\begin{eqnarray}
N=\int{\frac{\dot\alpha}{\dot\chi}~d\chi}\simeq \int{\frac{3(1+Q)\dot\alpha^2 f^2}{\mu^4\sin{\frac{\chi}{f}}}}~d\chi=f^2\int^{\pi}_{\frac{\chi_i}{f}}{\frac{(1+Q)(1+\cos{X})}{\sin{X}}}~dX\, ,
\label{enw}
\end{eqnarray}
where we defined $X=\chi/f$.
If we take $X=10^{-2}$ as an initial value, the number of e-folding is given by $N\sim 10 \ f^2 (1+Q)$. For example, $1+Q\sim f^{-2}\mathcal{O}(10)$ gives $N\sim \mathcal{O} (100)$. Hence, $Q$ should be large $Q\gg1$ for a natural value of the decay constant $f<1$.

\section{Chromo-natural warm inflation}
As shown in the previous sections, both chromo-natural inflation and minimal warm inflation assume that the interaction between inflaton and non-abeliean gauge fields through Chern-Simons term. Therefore, it is natural to ask if chromo-natural inflation and minimal warm inflation could be compatible or not.
The purpose of this section is to show the existence of chromo-natural warm inflation.

\subsection{Slow-roll equations of axion inflation
with Chern-Simons term}
We consider the same Lagrangian with chromo-natural inflation
\begin{eqnarray}
S=\int d^4 x \sqrt{-G}
\left[
\frac{1}{2} R -\frac{1}{2} \partial_\mu \chi \partial^\mu \chi -V(\chi)-\frac{1}{4}F^a_{\mu\nu}F_a^{\mu\nu}-\frac{g^2 \lambda}{8 f}\chi \epsilon^{\mu\nu\lambda\rho}F^a_{\mu\nu} F^a_{\lambda\rho}
\right]\, ,
\label{action3}
\end{eqnarray}
where the axion potential $V(\chi)$ is given in Eq.~\eqref{potential}.
We take into account the energy density of radiation in Eqs.~\eqref{EOMCNI1} and \eqref{EOMCNI2} as
\begin{align}
&\dot\alpha^2=\frac{1}{3}\left[\frac{1}{2}\dot\chi^2+\mu^4\left(1+\cos{\frac{\chi}{f}}\right)+\frac{3}{2}\dot\psi^2+3\dot\alpha\dot\psi\psi+\frac{3}{2}\dot\alpha^2\psi^2+\frac{3}{2}g^2\psi^4 +\rho_R\right] \, ,
\label{EOMCNwI1}\\
&\ddot\alpha=-\frac{1}{2}\dot\chi^2-\dot\psi^2 -2\dot\alpha\dot\psi\psi
-\dot\alpha^2\psi^2-g^2\psi^4-\frac{2}{3}\rho_R\, ,
\label{EOMCNwI2}
\end{align}
and Eqs.~\eqref{EOMCNI3} and \eqref{EOMCNI4} read
\begin{align}
&\ddot\chi+3\dot\alpha (1+Q)\dot\chi-\frac{\mu^4}{f}\sin{\frac{\chi}{f}}=-3\frac{g^3\lambda}{f}\psi^2(\dot\psi+\dot\alpha \psi) \, ,
\label{EOMCNwI3}\\
&\ddot\psi+3\dot\alpha\dot\psi+\left[-\frac{1}{6}\dot\chi^2+\frac{2}{3}\mu^4\left(1+\cos{\frac{\chi}{f}}\right)\right]\psi+2g^2\psi^3-\frac{g^3\lambda}{f}\psi^2\dot\chi=0 \, ,
\label{EOMCNwI4}
\end{align}
where we added the dissipation term only for the axion equation of motion.
It is convenient to define $\rho_{\chi}$  and $ \rho_{\psi}$ as 
\begin{align}
&\rho_{\chi}=\mu^4\left(1+\cos{\frac{\chi}{f}}\right) \, ,
\label{Ex}\\
&\rho_{\psi}=\frac{3}{2}\dot\alpha^2\psi^2+\frac{3}{2}g^2\psi^4 \, .
\label{Ep}
\end{align}
The energy conservation implies
\begin{align}
\dot{\rho}_{\rm R}+4\dot\alpha\rho_{\rm R}=3\dot\alpha Q \dot{\chi}^2 \, .
\label{CNwIeom3}
\end{align}

In the slow-roll approximation, we impose the condition $|\ddot\chi|\ll \dot\alpha(1+Q)|\dot\chi|$, $|\ddot\psi|\ll \dot\alpha|\dot\psi|$, $\dot\rho_{\rm R}\ll \dot\alpha\rho_{\rm R}$ and the inflaton potential is dominant in Eqs.~\eqref{EOMCNwI1}, \eqref{EOMCNwI3}, \eqref{EOMCNwI4} and \eqref{CNwIeom3}.
The equations are simplified as follows
\begin{align}
&\dot\alpha^2=\frac{\mu^4}{3}\left(1+\cos{\frac{\chi}{f}}\right) \, ,
\label{EOMCNwIs1}\\
&3\dot\alpha(1+Q) \dot\chi=\frac{\mu^4}{f}\sin{\frac{\chi}{f}}-3\sigma\dot\alpha(\dot\psi+\dot\alpha \psi) \, ,
\label{EOMCNwIs3}\\
&3\dot\alpha\dot\psi=-2\left(1+\frac{\sigma^2}{\gamma}\right)\dot\alpha^2\psi+\sigma\dot\alpha\dot\chi \, ,
\label{EOMCNwIs4}\\
&\rho_{\rm R}=\frac{3}{4} Q\dot{\chi}^2 \, .
\label{CNwIeoms3} 
\end{align}
Note that we defined $\sigma\equiv\frac{g^3\lambda\psi^2}{f\dot\alpha}$ and $\gamma\equiv\frac{g^4\lambda^2\psi^2}{f^2}$. 
The consistency conditions for the slow-roll approximation can be derived as follows. From the condition $\dot\rho_{\rm R}\ll \dot\alpha\rho_{\rm R}$, 
we obtain
\begin{align}
&\epsilon_{\rm v}\equiv\frac{1}{2(1+Q)\left(1+\frac{\sigma^2}{3(1+Q)}\right)}\left(\frac{V'(\chi)}{V(\chi)}\right)^2\ll 1 \,, \label{cnwsc1}\\
&\eta_{\rm v}\equiv\frac{1}{(1+Q)\left(1+\frac{\sigma^2}{3(1+Q)}\right)}\frac{V''(\chi)}{V(\chi)}\ll 1 \,,\label{cnwsc2} \\
&\beta\equiv\frac{1}{1+Q}\frac{Q'\,V'(\chi)}{Q\,V(\chi)}\ll 1~\label{cnwsc3}\,.
\end{align}
We can see that the slow-roll conditions \eqref{cnsc1}, \eqref{cnsc2} and \eqref{sc4} $\sim$ \eqref{sc6} are included as special cases of Eqs.~\eqref{cnwsc1}, \eqref{cnwsc2} and \eqref{cnwsc3}. The requirement $|\ddot\psi|\ll \dot\alpha|\dot\psi|$ does not imply any additional condition. From the condition $|\ddot\chi|\ll \dot\alpha(1+Q)|\dot\chi|$, we can see that the slow-roll condition \eqref{cnsc3} is updated to
\begin{align}
&\delta\equiv\frac{1}{\left(1+\frac{\sigma^2}{3(1+Q)}\right)}\frac{\sigma^2}{\gamma}\frac{1}{(1+Q)}\ll 1 \,. 
\label{cnwsc4}
\end{align}
Note that these slow-roll conditions can be satisfied, when all the parameter $\sigma$, $\gamma$ and $Q$ are larger than one.

After diagonalizing  Eqs.~\eqref{EOMCNwIs3} and \eqref{EOMCNwIs4}, the equations of the axion and the gauge field are given by 
\begin{align}
&\dot\alpha\dot\chi=\frac{1}{9\left(1+\frac{\sigma^2}{3(1+Q)}\right)}\left[\frac{3\frac{\mu^4}{f}\sin{\frac{\chi}{f}}}{1+Q}-\left(1-\frac{2\sigma^2}{\gamma}\right)\frac{\mu^4\left(1+\cos{\frac{\chi}{f}}\right)\sigma\psi}{1+Q}\right] \, ,
\label{EOMCNwIchi}\\
&\dot\alpha\dot\psi=\frac{1}{9\left(1+\frac{\sigma^2}{3(1+Q)}\right)}\left[\frac{\sigma \frac{\mu^4}{f}\sin{\frac{\chi}{f}}}{1+Q}-\left(\frac{\sigma^2}{1+Q}+2+\frac{2\sigma^2}{\gamma}\right)\mu^4\left(1+\cos{\frac{\chi}{f}}\right)\psi\right] \, .
\label{EOMCNwIspsi}
\end{align}

\subsection{Effect of dissipation on chromo-natural inflation }\label{subsection1}
It is possible to cast Eq.~\eqref{EOMCNwIspsi} into the form $\dot\alpha\dot\psi\simeq-V'(\psi)_{\rm eff}$ where we defined the effective potential 
 \begin{align}
V'(\psi)_{\rm eff}=\frac{\mu^4}{3\left(1+\frac{3(1+Q)}{\sigma^2}\right)}\left[-\frac{ \sin{\frac{\chi}{f}}}{f\sigma }+\left(1+\frac{2(1+Q)}{\sigma^2}+\frac{2(1+Q)}{\gamma}\right) \left(1+\cos{\frac{\chi}{f}}\right)\psi\right] \, .
\label{effectivepotential}
\end{align}
From Eq.~\eqref{effectivepotential}, we obtain the fixed point
\begin{align}
&\psi_{\rm fixed} = \left(\frac{\mu^2\sin{\frac{\chi}{f}}}{\sqrt{3}g^3\lambda \sqrt{1+\cos{\frac{\chi}{f}}}\left(1+\frac{2(1+Q)}{\sigma^2}+\frac{2(1+Q)}{\gamma}\right)}\right)^{1/3} \, .
\label{psimin4}
\end{align}
The second and third terms in the denominator represent corrections to the attractor solution of chromo-natural inflation due to the effect of dissipation $Q$. This formal solution shows that  chromo-natural inflation can be realized when $Q$ is much smaller than  $\sigma^2$ and $\gamma$. However, once $Q$ becomes comparable to $\sigma^2$ and $\gamma$, we would expect warm inflation 
commences. What we would like to clarify is if chromo-natural and warm inflation 
coexist or not.

Let us take a look at the following expressions
\begin{align}
    \sigma=\frac{g^3 \lambda \psi^2}{f \dot\alpha}~,~~\gamma=\frac{g^4 \lambda^2 \psi^2}{f^2}~,~~Q =\kappa g^{10} T^3\frac{\lambda^2}{3\dot\alpha f^2}\,.
\end{align}
 Once the gauge field is captured at the minimum of the effective potential,  $\sigma^2$, $\gamma$ and $Q$ are determined by the parameters $g$, $f$, $\lambda$, $\mu$ and $\kappa$ and the field $\chi$. Now, we change the parameter $\kappa$ which controls the effect of dissipation, and fix the other parameters
 and the field $\chi$. For $\kappa=0$, the solution reduces to that of chromo-natural inflation in section \ref{section2}. As we increase $\kappa$, the effect of dissipation becomes larger. Eventually, we see 
 chromo-natural warm inflation can be realized. 
 We will find that chromo-natural inflation ceases to occur for larger $\kappa$.

 From now on, we fix the parameters $\{g,f,\lambda,\mu\}=\{0.2,~0.01,~9\times 10^{5},~3\times10^{-4}\}$. 
For $\kappa =0$, we have an attractor solution 
\begin{align}
\psi_{\rm min} &=  \left(\frac{\mu^2\sin{\frac{\chi}{f}}}{\sqrt{3}g^3\lambda \sqrt{1+\cos{\frac{\chi}{f}}}}\right)^{1/3} 
\simeq \left(\frac{\mu^2}{\sqrt{3}g^3\lambda}\right)^{1/3}
\notag\\
& \simeq 0.0002
\left(\frac{0.2}{g}\right)
\left(\frac{9\times 10^5}{\lambda}\right)^{\frac{1}{3}}
\left(\frac{\mu}{3\times10^{-4}}\right)^{\frac{2}{3}}
\, ,
\label{psi1}
\end{align}
where we assumed the $\chi$ dependent part gives 
${\cal O} ( 1 )$
contribution.
Then, $\sigma$ and $\gamma$ would be given by 
\begin{align}
\sigma\simeq 10^5\left(\frac{g}{0.2}\right)\left(\frac{10^{-2}}{f}\right)\left(\frac{\lambda}{9\times 10^5}\right)^{\frac{1}{3}}\left(\frac{3\times10^{-4}}{\mu}\right)^{\frac{2}{3}} \, ,
\end{align}
and
\begin{align}
\gamma\simeq 10^{6}
\left(\frac{g}{0.2}\right)^2\left(\frac{10^{-2}}{f}\right)^{2}\left(\frac{\lambda}{9\times 10^5}\right)^{\frac{4}{3}}\left(\frac{\mu}{3\times10^{-4}}\right)^{\frac{4}{3}}\, .
\end{align}
 Since $\sigma\gg1$ and $\gamma\gg1$, the slow-roll condition of chromo-natural inflation are satisfied. Notice that, for these parameters, $\sigma^2\gg\gamma$ is satisfied. 
 
Let us start to increase $\kappa$ gradually.
As long as the effect of dissipation can be negligible, $\sigma^2\gg Q$ and $\gamma\gg Q$,  the fixed point is approximated by that of chromo-natural inflation $\psi_{\rm min}$. 
 The slow-roll conditions, from \eqref{cnwsc1} to \eqref{cnwsc4}, are reduced into those of chromo-natural inflation $\sigma\gg1$ and $\gamma\gg1$. 
  Using Eqs.~\eqref{dissipation}, \eqref{rho} and \eqref{CNwIeoms3},
  we obtain 
\begin{eqnarray}
    T= \frac{3}{40\pi^2} \kappa g^{10} 
    \frac{\lambda^2}{\dot{\alpha} f^2} \dot{\chi}^2 \ ,
    \label{T}
\end{eqnarray}
where we chose $g_i=100$ in Eq.~\eqref{rho}.
Since the inequality $\sigma^2\gg\gamma$ holds and $\dot\psi\simeq0$ is satisfied, Eq.~\eqref{EOMCNwIs4}
yields
\begin{eqnarray}
    \dot{\chi}=  \frac{2f}{g \lambda} \psi_{\rm min} \, .
    \label{chi}
\end{eqnarray}
 Substituting  Eqs.~\eqref{psi1} and \eqref{chi} into Eq.~\eqref{T},
 we can analytically obtain the ratio of temperature to the Hubble parameter
 as 
\begin{align}
\frac{T}{\dot\alpha} = \frac{1}{30}\kappa g^6 \left(\frac{ g^2 \psi_{\rm min}^2}{\dot\alpha^2}\right)\simeq \kappa~ \left(\frac{g}{0.2}\right)^6\left(\frac{3\times 10^{-4}}{\mu}\right)^{\frac{8}{3}}\left(\frac{9\times 10^5}{\lambda}\right)^{\frac{2}{3}} \, .
\label{t1}
\end{align}
Thus, for $\kappa > 1$, we see that warm inflation is realized. 
Using Eq.~\eqref{t1}, we can evaluate the dissipation parameter \eqref{dissipation} as
\begin{align}
Q&=\kappa g^{10} \frac{\lambda^2}{3\dot\alpha f^2}T^3 \simeq 10^{-7}\kappa^4\left(\frac{g}{0.2}\right)^{28}\left(\frac{3\times 10^{-4}}{\mu}\right)^4\left(\frac{10^{-2}}{f}\right)^2 \,.
\label{Q1}
\end{align}
We can also deduce the following expressions
\begin{align}
\frac{Q}{\sigma^2} \simeq 10^{-18} \kappa^4\left(\frac{g}{0.2}\right)^{26}\left(\frac{3\times 10^{-4}}{\mu}\right)^{\frac{8}{3}}\left(\frac{9\times 10^5}{\lambda}\right)^{\frac{2}{3}} \, ,\label{QS1}
\end{align}
and
\begin{align}
\frac{Q}{\gamma} \simeq 10^{-12}\kappa^4\left(\frac{g}{0.2}\right)^{26}\left(\frac{3\times 10^{-4}}{\mu}\right)^{\frac{16}{3}}\left(\frac{9\times 10^5}{\lambda}\right)^{\frac{4}{3}} \,  .
\label{QG1}
\end{align}
From the above results,  we see the current assumptions $\sigma^2\gg Q$ and $\gamma\gg Q$ hold  as long as $\kappa$ satisfies the condition
$\kappa\ll 10^3$. 
For $1<\kappa$, we see the temperature could exceed the Hubble parameter, which is a characteristic feature of warm inflation.  In particular, for $10^{1.75} <\kappa\ll 10^3$,
we obtain the strong warm inflation $Q>1$.
Thus, chromo-natural warm inflation occurs for the parameter region $1<\kappa \ll 10^3$.
Note that these conditions change depending on the choice of other parameters. For example, when we consider $g\sim\mathcal{O}(10^{-2})$, the condition for warm inflation to occur becomes  $10^6<\kappa\ll10^9$.

From Eqs.\eqref{chi} and \eqref{Q1}, we obtain the radiation energy density as
\begin{align}
    \rho_R =\frac{3}{4}Q\dot\chi^2
    \simeq
    10^{-5}~\frac{\kappa^{4} g^{32}}{  \dot\alpha^{4} }\psi_{\rm min}^8\simeq10^{-28}\kappa^4 \left(\frac{g}{0.2}\right)^{24}\left(\frac{3\times 10^{-4}}{\mu}\right)^{\frac{8}{3}}\left(\frac{9\times 10^{5}}{\lambda}\right)^{\frac{8}{3}} \ .
    \label{rhoR}
\end{align}
where we used the relation $\sigma^2\gg\gamma$ in the last equality. 
The energy density of the gauge field can be obtained by substituting the fixed point \eqref{psi1} into Eq.~\eqref{Ep} as
\begin{align}
    \rho_\psi = \frac{3}{2}g^2 \psi_{\rm min}^4 \simeq 10^{-16}\left(\frac{\mu}{3\times 10^{-4}}\right)^{\frac{8}{3}}\left(\frac{0.2}{g}\right)^{2}\left(\frac{9\times 10^{5}}{\lambda}\right)^{\frac{4}{3}}.
\end{align}
We can also obtain the number of e-folding using Eq.~\eqref{ec}. For the current parameters, 
 we obtain  $\beta=\lambda^{1/3}\mu^{4/3}$ is $\mathcal{O}(10^{-2})$, and $N\simeq g^2\lambda~\mathcal{O}(10^{-2})\sim \mathcal{O}(100)$. 

\begin{figure}[H]
\centering
 \includegraphics[width=10.5cm]{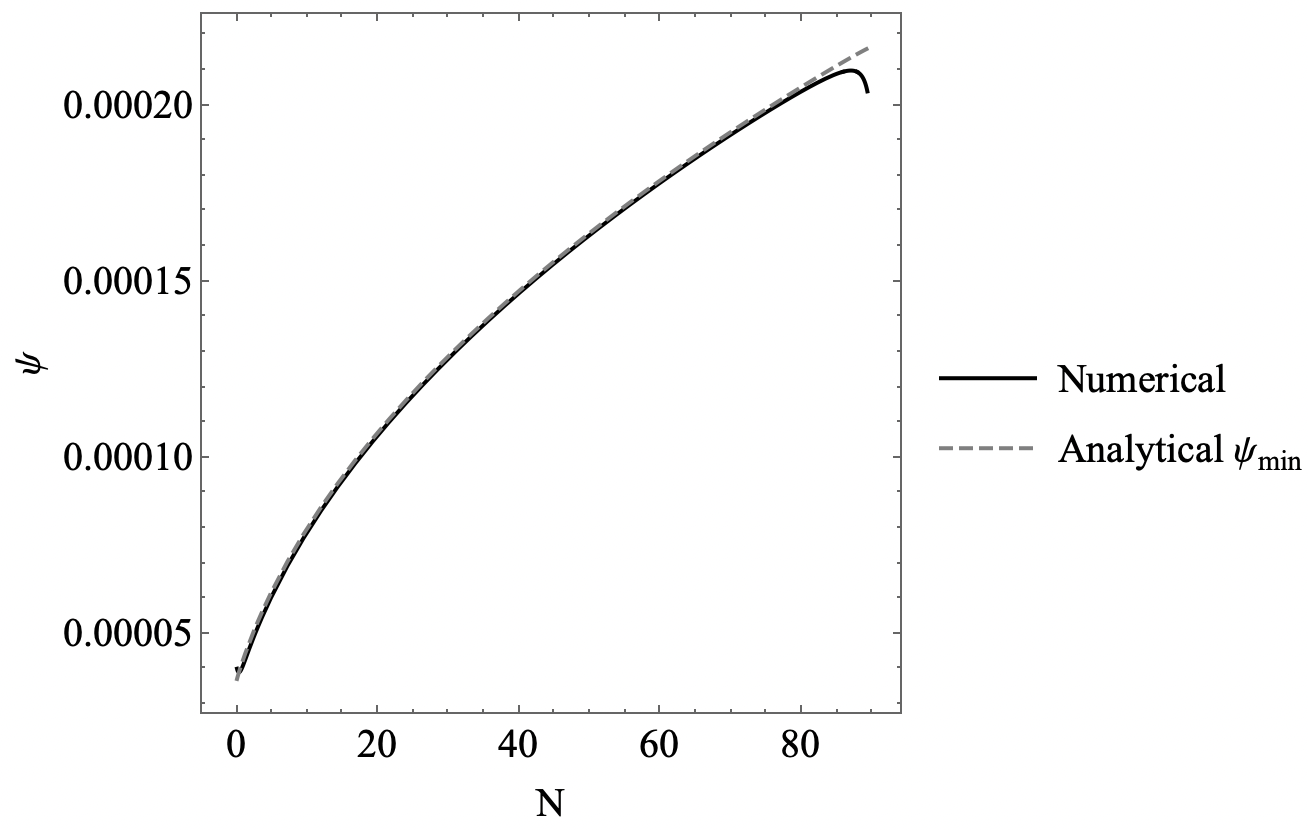}
 \renewcommand{\baselinestretch}{3}
 \caption{The solid curve represents the time evolution of the gauge field numerically obtained with $\kappa =10^2$. The dashed curve represents the analytical solution $\psi_{\rm min}$.}
\label{Fig1}
\end{figure}

\begin{figure}[H]
\centering
 \includegraphics[width=9cm]{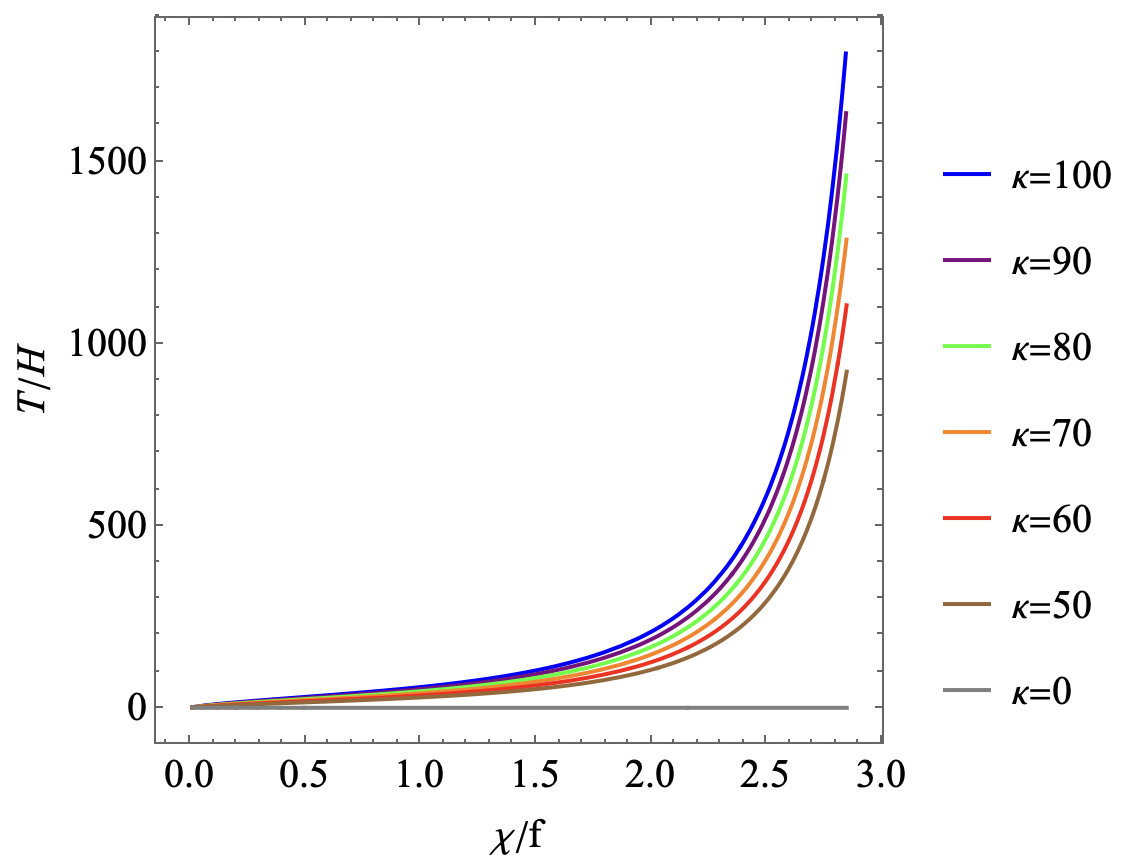}
 \renewcommand{\baselinestretch}{3}
 \caption{The horizontal axis represents ${\chi}/{f}$. Inflation ends when $\chi/f$ reaches $\pi$. The graph shows that larger $\kappa$ gives the larger temperature. We used the parameters: $g=0.2,~f=0.01, ~\lambda=9\times 10^{5},~ \mu = 3\times10^{-4}$ and the initial conditions: $ T_0 = 10^{-7},~\chi_0 = 10^{-4}~,~ v_{\chi_0} = 0,~ \psi_0 = 4\times10^{-5},~ v_{\psi_0} = 0$.}
\label{Fig3}
\end{figure}

To check the analytical results, we numerically solved basic Eqs.~\eqref{EOMCNwI3}, \eqref{EOMCNwI4} and \eqref{CNwIeom3}  with the parameter set $\{\kappa,g,f,\lambda,\mu\}=\{10^2,~2\times 10^{-1},~10^{-2},~9\times 10^{5},~3\times10^{-4}\}$.
 In Fig.~\ref{Fig1}, we can see that the gauge field value has settled down into the potential minimum $\psi_{\rm min}$, indicating that the slow-roll conditions are satisfied during inflation.  
  We also show $T/H$ for various $\kappa$  in Fig.~\ref{Fig3}. As we increase $\kappa$, we see the temperature gradually increases.

In this subsection, we have analytically investigate the chromo-natural warm inflation.
Although we have fixed the parameters other than $\kappa$, it is easy to change other parameters and to see what kind of inflation is realized. In particular, we should note that 
warm inflation is highly sensitive to the gauge coupling constant $g$ as one can see from 
Eqs.~\eqref{t1}, \eqref{Q1} and \eqref{rhoR}.

\subsection{The existence of chromo-natural warm inflation}\label{subsection2}
If we increase $\kappa$, the condition $\gamma\gg Q$ will be eventually violated, namely,   the effect of dissipation $Q$ cannot be negligible. According to Eq.~\eqref{psimin4}, the gauge field would become smaller as we increase $\kappa$. Hence, the slow-roll conditions of chromo-natural inflation, $\sigma\gg1$ and $\gamma\gg1$, will be violated in the end. In order to see the behavior, we will analyze the $\kappa$ dependence of the gauge field using the slow-roll Eqs.~ \eqref{EOMCNwIchi} and \eqref{EOMCNwIspsi}.

The temperature can be determined by the time derivative of axion $\dot\chi$ as in Eq.~\eqref{T}.  The slow-roll equation of the axion field gives  
\begin{align}
\dot\chi=\frac{\sqrt{3}}{9\left(1+Q+\frac{\sigma^2}{3}\right)}\left\{\frac{3\mu^2}{f}\frac{\sin{\frac{\chi}{f}}}{\sqrt{1+\cos{\frac{\chi}{f}}}}-\left(1-\frac{2\sigma^2}{\gamma}\right) \mu^2\sqrt{1+\cos{\frac{\chi}{f}}} \sigma\psi\right\} \, ,
\label{EOMCNwIchi2}
\end{align}
By substituting $\dot\chi$ into the temperature \eqref{T}, we obtain 
\begin{eqnarray}
    T= \frac{3\kappa g^{10}}{40\pi^2}  
    \frac{\lambda^2}{\dot{\alpha} f^2} \left[\frac{\sqrt{3}}{9\left(1+Q+\frac{\sigma^2}{3}\right)}\left\{\frac{3\mu^2}{f}\frac{\sin{\frac{\chi}{f}}}{\sqrt{1+\cos{\frac{\chi}{f}}}}-\left(1-\frac{2\sigma^2}{\gamma}\right) \mu^2\sqrt{1+\cos{\frac{\chi}{f}}} \sigma\psi\right\}\right]^2 
     ,  \quad
    \label{Ta}
\end{eqnarray}
Moreover, since the gauge field is settled down into the minimum of the effective potential, $\dot\psi\simeq 0$
will be realized. Thus, 
 Eq.~\eqref{EOMCNwIspsi} implies
\begin{align}
1+Q=\frac{\sigma^2}{2}\left(1+\frac{\sigma^2}{\gamma}\right)^{-1}  \left\{ \frac{1}{f\sigma \psi}\frac{ \sin{\frac{\chi}
{f}}}{1+\cos{\frac{\chi}{f}} }-1 \right\}
\, .
\label{q1}
\end{align}
By using Eq.~\eqref{q1} in Eq.~\eqref{Ta}, the temperature $T$ can be written as a function of $\psi$. On the other hand, using the expression of $Q$ \eqref{dissipation} in Eq.~\eqref{q1}, the temperature $T$ can be also written by $\psi$ as 
\begin{align}
T=\left(\frac{\sqrt{3}\mu^2 f^2\sigma^2}{2\kappa g^{10} \lambda^2}\right)^{\frac{1}{3}}\left[\left(1+\frac{\sigma^2}{\gamma}\right)^{-1}
\left\{\frac{1}{f\sigma \psi}\frac{ \sin{\frac{\chi}
{f}}}{\left(1+\cos{\frac{\chi}{f}} \right)}-1\right\}-\frac{2}{\sigma^2}\right]^{\frac{1}{3}}\, .
\label{Tb}
\end{align}
The attractor solutions of slow-roll inflation could be obtained as the intersection of \eqref{Ta} and \eqref{Tb}. To make the analysis precise, we fix the parameters $g$, $f$, $\lambda$, $\mu$.  Thus, the attractor of the gauge field and the temperature can be determined by the axion field $\chi$ once $\kappa$ is determined. 
During inflation, $\chi$ is almost constant.
Hence, we can take a representative value. 

Now, we will take a look at the $\kappa$ dependence of $\psi$, $\sigma$, $\gamma$, $\rm {T/H}$, $Q$ and $\dot\chi$.
 In the numerical calculations, we use the parameters: $g=0.2,~f=0.01, ~\lambda=9\times 10^{5}, ~ \mu = 3\times10^{-4}$. We also choose a reference point $\chi=0.01$ for the axion field.
\begin{figure}[H]
\centering
 \includegraphics[width=10cm]{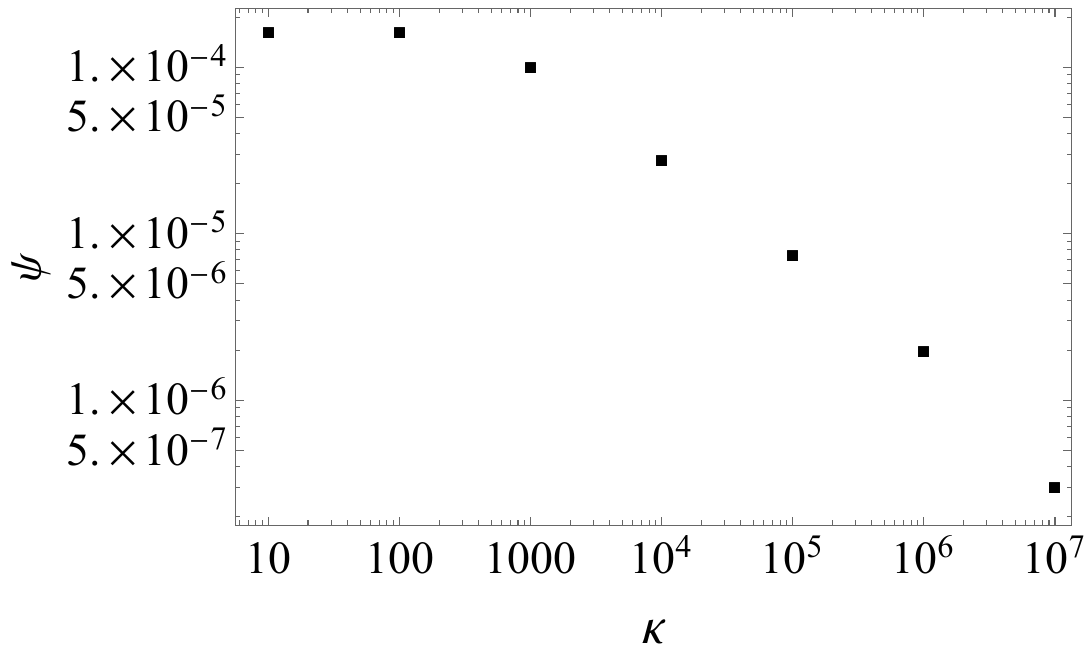}
 \renewcommand{\baselinestretch}{3}
 \caption{The gauge field $\psi$ is depicted for various $\kappa$. The gauge field does not depend on $\kappa$ for $\kappa <10^2$.
 For $\kappa >10^2$, the gauge field  gradually decreases as  $1/\sqrt{\kappa}$. }
\label{Fig4-3-1}
\end{figure}

\begin{figure}[H]
\centering
 \includegraphics[width=9.5cm]{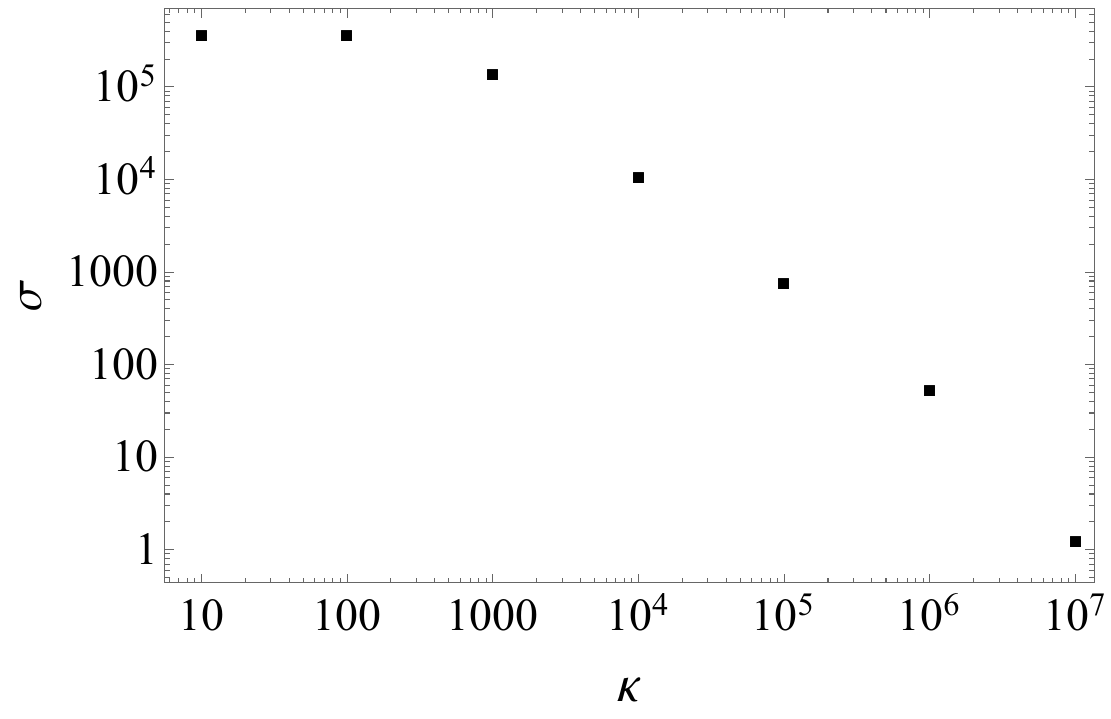}
 \renewcommand{\baselinestretch}{3}
 \caption{The variable $\sigma$ is depicted for various $\kappa$. Since $\sigma\gg1$  is the slow-roll condition for chromo-natural inflation,  we see that chromo-natural inflation ceases to occur around $\kappa=10^7$. }
\label{Fig4-3-2}
\end{figure}

\begin{figure}[H]
\centering
 \includegraphics[width=9.5cm]{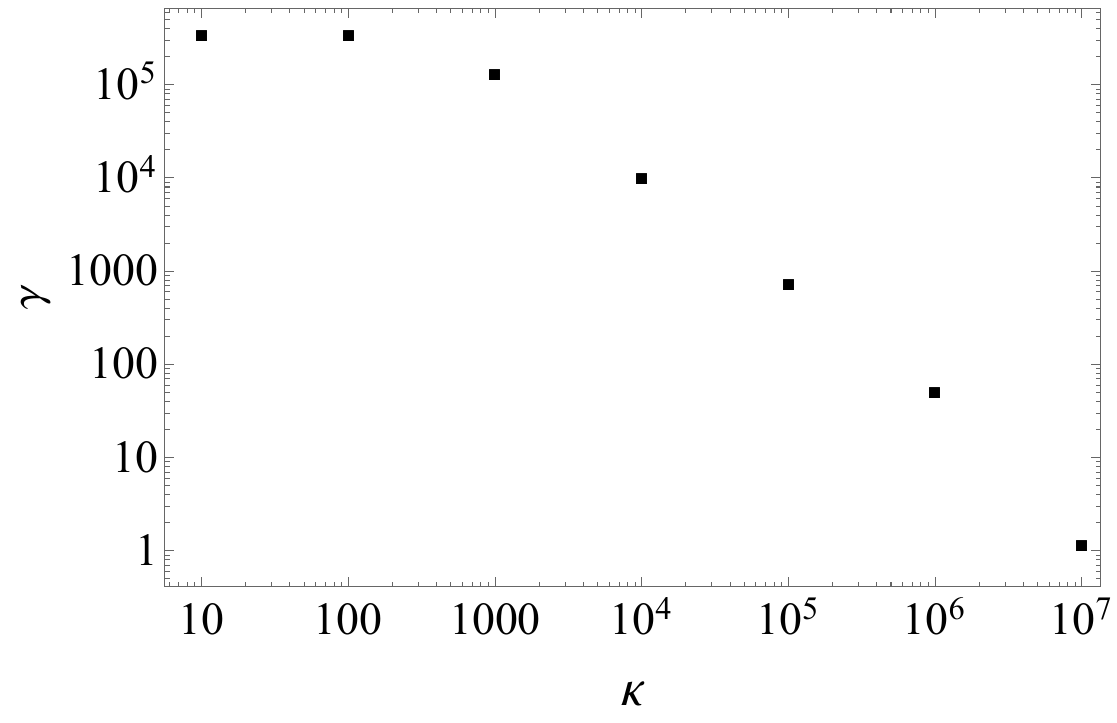}
 \renewcommand{\baselinestretch}{3}
 \caption{We plotted $\gamma$ for various $\kappa$. Since $\gamma\gg1$ is the slow-roll condition for chromo-natural inflation, again we see that chromo-natural inflation ceases to occur around  $\kappa=10^7$. }
\label{Fig4-3-3}
\end{figure}

Since we know the behavior of solutions up to $\kappa=10$ from the analysis in the previous subsection \ref{subsection1}, we will start with $\kappa=10$.
In Fig.~\ref{Fig4-3-1}, we plotted the gauge field for various $\kappa$. 
In Fig.~\ref{Fig4-3-2} and Fig.~\ref{Fig4-3-3},
the parameters $\sigma$ and $\gamma$ which characterize the slow roll chromo-natural inflation are depicted. 
Up to  $\kappa=10^2$, $\psi$, $\sigma$ and  $\gamma$ do not change as  shown in Fig.~\ref{Fig4-3-1}, Fig.~\ref{Fig4-3-2} and Fig.~\ref{Fig4-3-3}, which is consistent with the results in the previous subsection \ref{subsection1}. In the parameter region  $\kappa\gtrsim10^3$, we see the behavior $\psi\propto 1/\sqrt{\kappa}$ due to the dissipation. As a consequence, $\sigma$ and $\gamma$ decrease linearly $1/\kappa$. Around $\kappa=10^7$, we can see that $\sigma$ and $\gamma$ become $\mathcal{O}(1)$ which means that the chromo-natural inflation ceases to occur. In fact, we numerically checked that the intersection point  disappears when we take $\kappa>10^7$. 

\begin{figure}[H]
\centering
 \includegraphics[width=9.5cm]{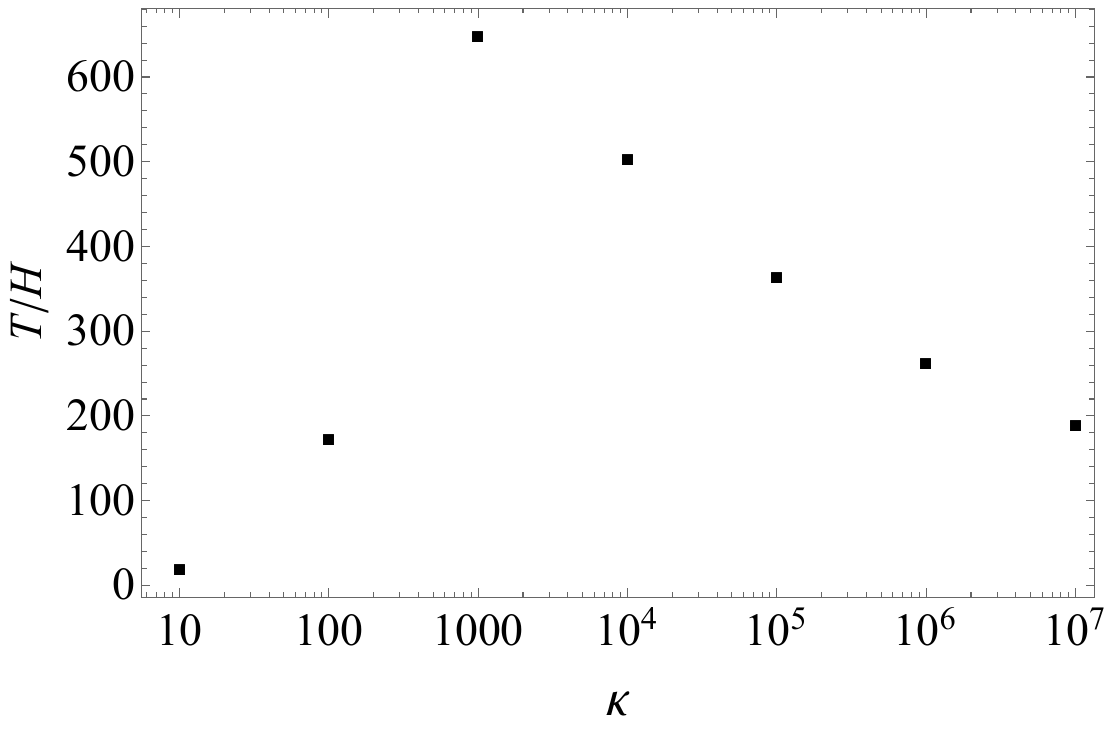}
 \renewcommand{\baselinestretch}{3}
 \caption{The ratio $T/H$ is depicted for various $\kappa$. Around $\kappa=10^3$, the ratio becomes maximum. }
\label{Fig4-3-4}
\end{figure}

In Fig.~\ref{Fig4-3-4}, we plotted the ratio of temperature to the Hubble parameter $T/H$ for various $\kappa$.
We find the ratio $T/H$ takes the maximum value $T/H \simeq 600$ around $\kappa=10^3$. Beyond $\kappa=10^3$, the ratio  $T/H$ decreases. 

\begin{figure}[H]
\centering
 \includegraphics[width=9.5cm]{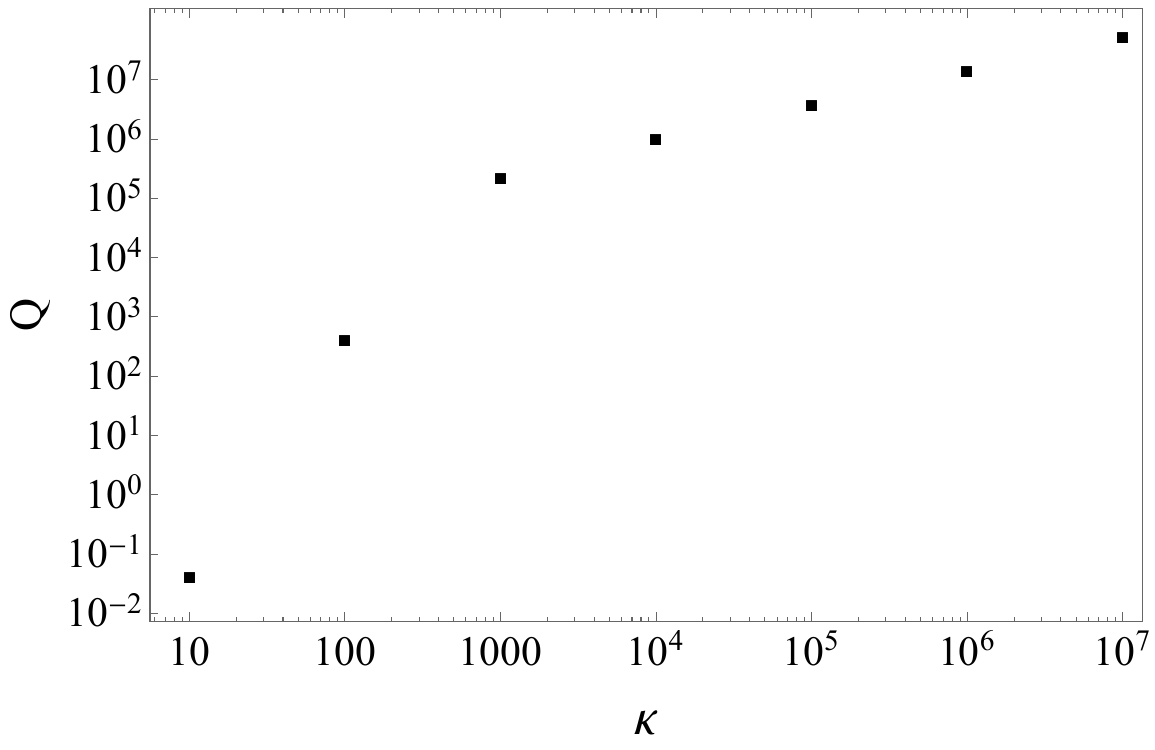}
 \renewcommand{\baselinestretch}{3}
 \caption{The dissipation parameter $Q$ is depicted for various $\kappa$. We see $Q$ rapidly increase up to $\kappa=10^3$, beyond which $Q$ increase slowly. }
\label{Fig4-3-5}
\end{figure}

 We also plotted the dissipation parameter $Q$ in Fig.~\ref{Fig4-3-5}. From Fig.~\ref{Fig4-3-5}, we see the dissipation parameter $Q$ rapidly increases up to $\kappa=10^3$, beyond which $Q$ increases slowly. 
 Note that the dissipation parameter $Q$ is always much larger than 1 for large $\kappa$.

\begin{figure}[H]
\centering
 \includegraphics[width=10cm]{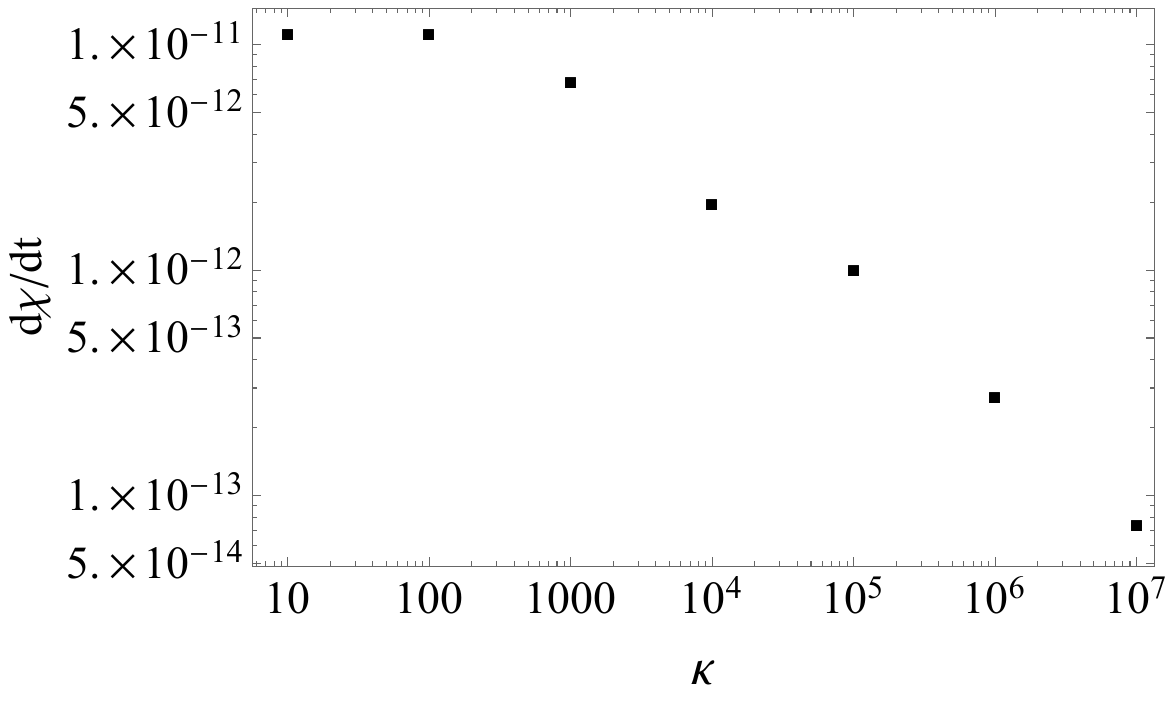}
 \renewcommand{\baselinestretch}{3}
 \caption{The time derivative of the axion field $\dot\chi$ is depicted for various $\kappa$.  From $\kappa=10^3$,  $\dot\chi$ decreases rapidly. }
\label{Fig4-3-6}
\end{figure}

Remarkably,  for $\kappa>10^3$, the temperature $T$ decreases in spite that $Q$ increases.  This is because $\dot\chi$ decreases faster than the increasing rate of $Q$.
 Note that 
the larger $\kappa$ enhances the dissipation
and as a consequence the friction becomes large.
That makes $\dot\chi$ small.
 To verify this, we plotted $\dot\chi$ in  Fig.~\ref{Fig4-3-6}.
 This is the reason why the ratio  $T/H$ starts to decrease at some point.

 To understand the behavior of the universe in the parameter region $\kappa>10^7$, we checked analytically and numerically the maximum value of $\kappa$ where warm inflation can occur. Beyond $\kappa=10^{7}$, since  the gauge field becomes trivial, the behavior of the axion field would return to that of warm inflation. The time derivative of the axion field $\dot\chi$ is given by taking $\psi=0$ in Eq.~\eqref{EOMCNwIchi2} as
\begin{align}
\dot\chi=\frac{1}{\sqrt{3}\left(1+Q \right)}\frac{\mu^2}{f}\frac{\sin{\frac{\chi}{f}}}{\sqrt{1+\cos{\frac{\chi}{f}}}} \, .
\label{Warm_chi}
\end{align}
 From Eqs.~\eqref{rho} and \eqref{CNwIeoms3}, the temperature is
 given by
\begin{align}
    T=\frac{1}{\kappa^{\frac{1}{7}}}\left\{\frac{9}{40\pi^2}\frac{\dot\alpha \mu^4}{  g^{10} \lambda^2 }\frac{ \sin^2{\frac{\chi}{f}}}{\left(1+\cos{\frac{\chi}{f}}\right)}\right\}^\frac{1}{7} \,.
    \label{Twarm}
\end{align}
Note that we assumed  $Q\gg1$ as shown in Fig.~\ref{Fig4-3-4}. 
As $\kappa$ is increased, we see the temperature decreases.
Thus, eventually we will reach the critical point $T=\dot\alpha$. Now,  assuming the parameter set: $g=0.2,~f=0.01, ~\lambda=9\times 10^{5}, ~ \mu = 3\times10^{-4}$ and the axion field $\chi=0.01$,
we can obtain $\kappa$ at the critical point
\begin{align}
    \kappa=\frac{3^5}{40\pi^2}\frac{ 1}{ g^{10} \lambda^2 \mu^8}\frac{\sin^2{\frac{\chi}{f}}}{\left(1+\cos{\frac{\chi}{f}}\right)^4}\simeq10^{22}\,.
\end{align}
Thus, for $\kappa >10^{22}$, the temperature is below the Hubble parameter, that is, we have cold inflation. In section~\ref{section3}, we obtained the number of e-folding \eqref{enw} which can be approximately written as $N\sim 10 f^2 Q$. Substituting \eqref{Twarm} into \eqref{dissipation}, we can derive the dissipation parameter $Q$ as
\begin{align}
    Q=\kappa^{\frac{4}{7}}\left\{\frac{3}{(40\pi^2)^3}\frac{ g^{40} \lambda^8 \mu^4}{ f^{14} }\frac{ \sin^6{\frac{\chi}{f}}}{\left(1+\cos{\frac{\chi}{f}}\right)^5}\right\}^\frac{1}{7} \,.
    \label{Q}
\end{align}
We see the number of e-folding increases as $N \propto \kappa^{\frac{4}{7}}$. Therefore, the universe looks like an eternal inflation for $\kappa > 10^{22}$.

In the previous subsection, we have analytically shown that
the chromo-natural warm inflation exists for $1<\kappa<10^3$.
In this subsection, we have numerically shown
that the chromo-natural warm inflation exists
in the parameter region $10<\kappa<10^7$. For $\kappa>10^7$, warm inflation is realized up to $\kappa<10^{22}$. If we consider much larger $\kappa>10^{22}$, almost eternal inflation is realized with the number of e-folding $N \propto \kappa^{\frac{4}{7}}$.
Thus, we can conclude that chromo-natural warm inflation exists for $1<\kappa<10^7$.

\section{Conclusion }
We studied axion inflation with the non-abelian Chern-Simons interaction term. In particular, we examined if chromo-natural inflation and minimal warm inflation  can occur simultaneously.  We showed that these inflation models can coexist.
Indeed, chromo-natural inflation could have a temperature $T>H$, which is a characteristic feature of warm inflation.

\begin{figure}[H]
\centering
 \includegraphics[width=14cm]{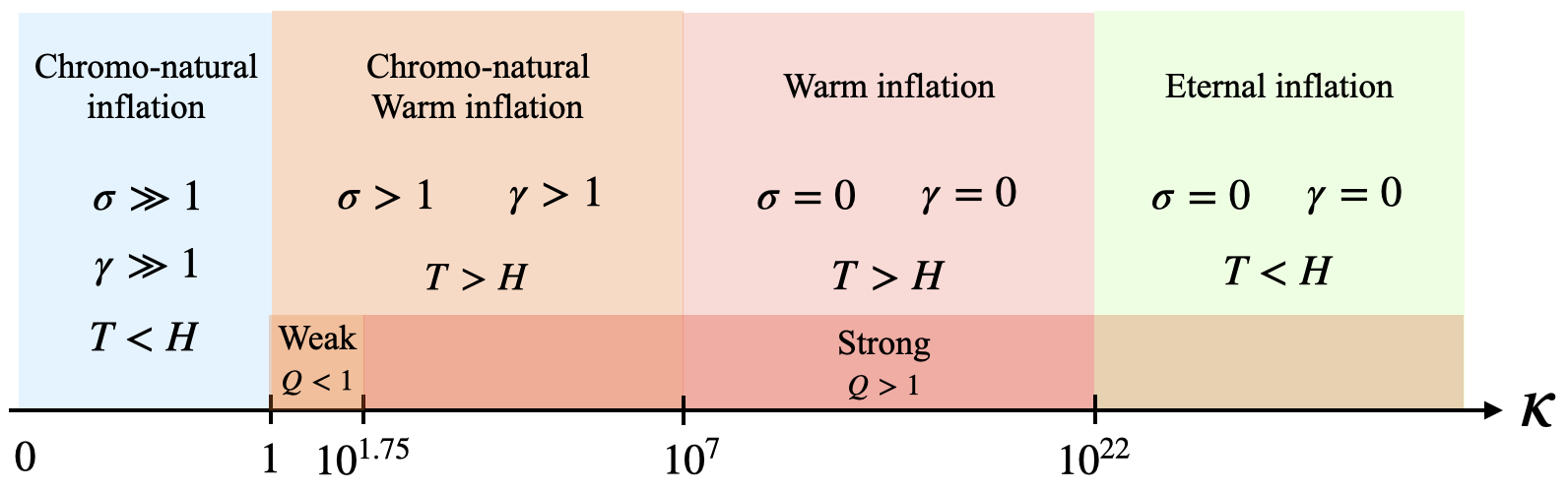}
 \renewcommand{\baselinestretch}{3}
 \caption{For the same parameters as those in Fig.~\ref{Fig4-3-1}, we depicted a one dimensional phase diagram of inflation. This diagram clearly shows that chromo-natural warm inflation exists for a wide range parameter including a natural one $\kappa= 10^2$.}
\label{Fig9}
\end{figure}

We analyzed the system in two ways.
First, starting from the chromo-natural inflation where $\kappa=0$, we gradually increased the parameter $\kappa$.
We analytically showed
chromo-natural warm inflation occurs
once $\kappa$ exceeds 1. In particular, when $\kappa$ is larger than $10^{1.75}$, the strong warm inflation appears.
Beyond $\kappa =10^3$, the approximation became invalid.
Hence, next, we numerically solved attractor equations.
We found gauge fields decreases as $\psi\propto 1/\sqrt{\kappa}$. Hence, $\sigma$ and $\gamma$ decreases linearly $1/\kappa$. Around $\kappa=10^7$, we see $\sigma$ and $\gamma$
become ${\cal O} (1)$. Namely, chromo-natural inflation ceases to occur there. 
We also found that the ratio of the temperature to the Hubble parameter $T/H$ always larger than $1$ in the parameter region $1<\kappa<10^7$.
Thus, we found that the chromo-natural warm inflation occurs in the parameter region $1<\kappa<10^7$.
For $\kappa >10^7$, we have warm inflation. 
If we further increase $\kappa$, since $\dot\chi$ decreases,  warm inflation ceases to occur around $\kappa =10^{22}$. For $\kappa >10^{22}$, 
inflation becomes cold and looks like an eternal inflation. 
We can summarize our findings in Fig.~\ref{Fig9}. 

In our analysis, we have always fixed the parameters other than $\kappa$.
However, our analytic expressions is useful for getting information of inflation with other values of parameters. 
For example, we can see the strong dependence of warm inflation on the gauge coupling constant $g$.
Moreover, our numerical analysis can be easily repeated
for other parameter cases. We just focused on the existence of chromo-natural warm inflation in this paper.

As we mentioned in the introduction, the counter example of the cosmic no-hair conjecture can  lead to interesting phenomenological consequences. In the present case, minimal warm inflation predicts the absence of primordial gravitational waves due to the low Hubble scale.  On the other hand, in chromo-natural inflation, the non-abelian gauge field contains the tensor modes interacting with the metric tensor modes, leading to the excess of primordial gravitational waves 
due to the tachyonic instability. However, in chromo-natural warm inflation, the growth of the gauge field would be suppressed. 
Hence, it might happen that tachyonic instability could produce an appropriate amount of gravitational waves which can be  observable. We will leave this issue for future work.

\section*{Acknowledgments}

J.\ S. was in part supported by JSPS KAKENHI Grant Numbers JP20H01902 and JP22H01220.

\printbibliography
\end{document}